\documentclass[aps,prl,reprint,groupedaddress]{revtex4-1}
\usepackage{graphicx}
\usepackage{amsmath}

\usepackage{graphicx}
\usepackage{dcolumn}
\usepackage{bm}
\usepackage{epstopdf}

\newcommand{\ext}{{\rm ext}}
\newcommand{\br}{{\bf r}}
\newcommand{\bx}{{\bf x}}
\newcommand{\ang}{{\rm ang}}

\newcommand{\isco}{{\rm rim}}
\newcommand{\muL}{{\mu L}}
\newcommand{\cE}{{\cal E}}
\newcommand{\eE}{{\rm ext}}
\newcommand{\LMC}{{\rm LMC}}
\newcommand{\ei}{{\rm ei}}
\newcommand{\mBD}{$\mu$BD}
\newcommand{\mBDs}{$\mu$BDs}
\newcommand{\jc}{{\rm jc}}
\newcommand{\bd}{{\rm bd}}
\newcommand{\gas}{{\rm gas}}

\newcommand{\myskip}[1]{}
\newcommand{\iso}{{\rm iso}}

\renewcommand{\d}{{\rm d}}

\newcommand{\BEQ}{\begin{eqnarray}}
\newcommand{\EEQ}{\end{eqnarray}}
\newcommand{\BEA}{\begin{eqnarray}}
\newcommand{\EEA}{\end{eqnarray}}
\newcommand{\nn}{\nonumber}

\newcommand{\cm}{{\rm cm}}

\newcommand{\K}{{\rm K}}

\newcommand{\kpc}{{\rm kpc}}

\newcommand{\half}{\frac{1}{2}}

\begin{document}

\title{A partially occulting MACHO-microlensing event in the Twin Quasar  Q0957+561}

\author{Theodorus Maria Nieuwenhuizen}

\address{ 
$^1$Institute for Theoretical Physics,  University of Amsterdam,  Science Park 904, 1090 GL  Amsterdam, The Netherlands \\
$^2$International Institute of Physics, UFRG,  Anel Vi\'ario da UFRN -- Lagoa Nova, Natal -- RN, 59064-741, Brazil}

\begin{abstract}
A doubly-peaked quasar microlensing event in the lensed Twin Quasar  Q0957+561 A,B (Colley and Schild 2003) is 
analysed within several lensing models.
In the most realistic model a lens resolves in image B the ellipse shaped, bright inner rim of the quasar's accretion disk,
intersecting it twice.
This lens weighs 0.5 Earth mass and is located inside the Galaxy, at 3 kpc distance.
During the passing, it partially occults the source, which allows to describe it as a primordial gas cloud 
of 1.4 Solar radius and 17 K temperature, in accordance with  the theory of Gravitational Hydrodynamics. 
Lensing by such objects against the Magellanic Clouds and Galactic centre  will also lead to occultation dips.
\end{abstract}

\maketitle



\section{Introduction}

Massive Astrophysical Compact Halo Objects, MACHOs, putative objects with mass between sublunar and solar, are the most natural
candidates for the Galaxy's missing  matter (dark matter), since they could solve the problem of the missing baryons:
while at the Galactic level 10\% of the cosmic budget can be attributed to known objects,
at cosmic scale 30\% of the baryons may be missing \cite{shull2012baryon}. (Dark matter of galaxy clusters
and at their scale must be of non-baryonic nature.)

But the case of MACHO dark matter has suffered a rough ride. Originally populations of MACHOs
were considered as the cause of all dark matter. The interpretation of brightness fluctuations in the Twin Quasar Q0957+561 A,B 
as caused by massive objects of a few Earth masses in its lensing galaxy supported the case \cite{schild1996microlensing}.
However, direct searches for microlensing of stars in the Magellanic clouds by MACHOs passing in front of them,
carried out by the EROS \cite{renault1998search,tisserand2007limits} and the MACHO  \cite{alcock1998eros,alcock2000macho} collaborations, 
turned up empty handed. Since then the mantra is: MACHO dark matter is ruled out.

Great was the surprise to detect  by microlensing against the Galactic centre an unbound or distant Jupiter-mass population,
which outnumbers the main-sequence stars by about a factor two \cite{microlensing2011unbound}. 
Further investigation shows that at least half of them are indeed free floating planets \cite{clanton2016constraining}.
The rate for free floating planet observation against the Galactic centre is estimated at dozens of Earth masses and over a thousand Jupiters 
per year \cite{ban2016microlensing}.
Furthermore, because the Galactic halo appears to be less heavy than it was assumed a decade ago, the conclusion of the MACHO team that MACHOs 
are ruled out as the main component of Galactic dark matter has been questioned \cite{hawkins2015new}.
The situation presently thus welcomes any positive or negative result.
We shall investigate another case of microlensing.

Planck has observed the Andromeda galaxy and found its dust temperature to have  a maximum of 22 K in the nucleus and decaying linearly to 14 K 
outside of the 10 kpc central ring, with the spiral rings extending to 26 kpc \cite{ade2015planck}.
Many cold ($T>15$ K) clouds are observed by Herschel, in particular in star forming regions, 
with possibly a colder core with $T$ down to 10 K \cite{juvela2011galactic,juvela2012galactic}.
These observations put forward a fundamental role for hydrogen clouds that get cooled towards the triple point at 13.8 K
and pass the first order transition line towards  the liquid and solid phase.

When a massive object  passes in front of a star, it acts as a gravitational lens that distorts and enhances the light.
In the Galaxy, typical distortions are micro arcseconds in size, hence the term microlensing.
But if the microlenseis a large, dense gas cloud, the light that comes to us may be absorbed on its far side, whence {\it occultation} takes place.

 A 15 K primordial H cloud of Earth mass with $\phi_{\rm He}=25\%$ weight in He,
has a large isothermal radius,  $R_\iso=GM_\oplus\mu m_N/2k_BT=2.85R_\odot$, where $\mu=1/(1-0.75 \phi_{\rm He})$ is the mean weight
and $m_N$ the nucleon mass. Given its huge column density $\sim10^{28}\cm^{-2}$, a light ray will be extinct when its impinges the cloud, 
up to the light extinction radius $R_\ext\gtrsim R_\iso$ where the column density has fallen to $10^{22}-10^{23}\cm^{-2}$. 
Such Earth--mass, Solar--size cold clouds may thus induce occultation in lensing events  towards the Magellanic clouds and in the Twin Quasar.
The latter case has provided data which we analyse first. 

In section 2 we recall properties of the Twin Quasar and the luminous inner ring of its accretion disk.
In section 3 we consider five models for the Colley-Schild 2003 lensing event.
The related  physical aspects are discussed in section 4. In section 5 we consider the
possibility of occultation in lensing of the LMC, while in section 6 we connect to the theory
of Gravitational Hydrodynamics. We  close with a  conclusion.

\section{On the Twin Quasar and its lensing} 

The intrinsic variations of quasars make it hard to distinguish their microlensing events, 
but lensed quasars may expose them because they will occur in one of the images, while the other image(s) can be taken as reference.
When data from the quasar images are at disposal, the knowledge of the time delay is then mandatory.
The QuOC-Around-The-Clock  consortium of 12 observatories around the globe continuously monitored for 10 days in January 2000 the A image 
of the quasar Q0957+561 and in March 2001 the B image, setting their time delay as $417.09\pm0.07$ day \cite{colley2003around}. 
Next, Colley and Schild \cite{colley2003rapid} (to be denoted as CS03) reconsider observations of 5 nights in 1994.9 (A image) and 1996.1 (B image) and adjust the
-- slowly changing -- time delay for that period to $417.07\pm0.07$ days.  
Such a small error allows to subtract the intensities in the images, which exhibits  that in the fifth night a clear lensing event occurs.
The authors connect it to a sublunar mass microlensein the lensing galaxy. However, such a light cloud should have evaporated
already  \cite{deRujula1992nature}.
We shall reinterpret the lensing event by a more precise modelling of the situation.

 \subsection{Specific properties}

Reverberation processes expose the luminous inner rim of the quasar accretion disk (shortly: the rim).
This sets its radius as $R_q=4.0\,10^{16}$ cm, where cosmology enters merely through the redshift \cite{schild2005accretion}.
The angular distance  $d_A(z_q)=1.123$ Gpc, following from the WMAP7 cosmology 
$\Omega_M = 0.1344h^{-2}$, $\Omega_B  = 0.022246h^{-2}$, $\Omega_\Lambda=1-\Omega_M$ \cite{hinshaw2013nine} and $h=0.74$ 
\cite{riess20113}, leads to an opening angle $\theta_q=R_q/d_A(z_q)=2.4\,\mu$as.

We shall be interested in this rim, because it will be involved in lensing, as we discuss below.
First we have to describe its basically elliptic shape on the sky. We take an observation frame with $z$-axis pointing to the heart of the quasar
and  the $x$-axis set by the lensing plane (quasar, lensing galaxy and the Sun), so that both quasar images lie on the $x$-axis\footnote{The text will be 
written from the point of view that the two quasar images lie on a straight line, with the lensing galaxy G1 also on the same line. Actually, the lens galaxy 
is slightly displaced from the line joining the two images, due to a small shear originating in the galaxy cluster containing G1. 
The small discrepancy that follows from this is unlikely to affect the present calculations.}.
Finally, the $y$-axis is perpendicular to the lensing plane.
In the quasar rest frame, the rim is a circle that can be parametrised as  $\vec R=R_q(\cos\phi,\sin\phi,0)$ with $-\pi<\phi<\pi$.
It is inclined over an angle $\alpha=55^\circ$ with respect to the $z$-axis \cite{schild2005accretion}. 
Hence in the frame where this rotation lies in the $y$-$z$ plane the rim has $3d$ position:
$ \vec{R}=R_q(\cos\phi,\sin\alpha\sin\phi,\cos\alpha\sin\phi)$,
and of this we observe the ($x,y$) components located on the $2d$ angular ellipse  $\theta_q(\cos\phi,\sin\alpha\sin\phi)$.
This vector makes an angle $\beta$ with respect to our $x$-axis, so that in the absence of the lensing galaxy we would observe the rim as an elllipse
$\bx(\phi)=\theta_q(\cos\beta\cos\phi+\sin\alpha \sin\beta\sin\phi, -\sin\beta\cos\phi+\sin\alpha\cos\beta\sin\phi)$  around the heart line to the quasar,
with $-\pi<\phi<\pi$.

By the lensing galaxy two images (A and B, or $\mp$) of the rim are created;
 their $\mu$as size can not be resolved optically, but we will see below that microlensing can achieve a resolution by 
 creating a specific profile with two peaks.
The lensing galaxy squeezes the components
(by factors $0<\mu_\pm^x<1$) in the lensing plane ($x$-direction)
and elongates them in the normal to the plane ($y$-direction), by  factors $\mu_+^y>1$ or $\mu_-^y<-1$ that we specify below.
($\mu_-^y<0$ refers to inversion of the picture.)
The quasar inner rim  is thus viewed around centres of the $\pm$ images at the angular positions

\BEQ \bx_\pm(\phi)= 
\theta_q\left(\mu_\pm^x(\cos\beta\cos\phi+\sin\alpha \sin\beta\sin\phi) \atop \mu_\pm^y(\sin\alpha\cos\beta\sin\phi -\sin\beta\cos\phi)\right),
\EEQ
where $-\pi<\phi<\pi$.
Lensing theory tells us that for a point source $S$ and a point lens $L$ of mass $M$ at distance $d_L$, the angular and physical Einstein radius read
\BEQ
\theta_E
=\sqrt{\frac{4GMd_{SL}}{c^2d_{S}d_L}}
,\qquad 
R_E
=\sqrt{\frac{4GMd_{SL}d_L}{c^2d_{S}}},
\EEQ
with $d_S$ the distance to $S$ and $d_{SL}$ the one from $S$ to $L$.
With the source at the angular position $\bx$ and, likewise,  the lens at $\br$, there occur two images at the angles 

\BEQ 
\theta_\pm=\half \theta_E(u\pm\sqrt{u^2+4}), \qquad u=\frac{\vert\bx-\br\vert}{\theta_E}.
\EEQ
The magnification factors in the lensing plane, $\mu_\pm^x=(\d\theta_\pm/\d u)/\theta_E$ and perpendicular to it, 
$\mu_\pm^y=\theta_\pm/u\theta_E$,  imply a total magnification   $\mu_\pm=\mu_\pm^x\mu_\pm^y$ equal to

\BEQ \label{mupm=}
\mu_\pm(u)=\frac{1\pm\mu(u)}{2},\qquad
\mu(u)=\frac{2+u^2}{u\sqrt{4+u^2}}.
\EEQ

For the Twin Quasar the A and B images are separated by $6''$. 
The A image has average apparent magnitude $\bar m_A=16.7$ and the brighter  B image $\bar m_B=16.5$.
When approximating the lensing galaxy as a point mass, this brings $\bar m_A-\bar m_B=2.5\log_{10}{\mu_+}/{\vert\mu_-\vert}$, 
implying the impact parameter $u=0.0921$ of the lensing galaxy. It follows that $\mu_+^x=0.523$, $\mu_+^y=11.4$ and $\mu_x^-=0.477$, $\mu_-^y=-10.4$, so that
the quasar intensity is magnified in the B image by $\mu_+=5.94$ and in the A image by $\mu_-=-4.94$.

A more detailed modelling of the lensing galaxy has appeared to be difficult, without a clear result up to now \cite{fadely2010improved}.
What enters in our analysis are the magnification factors $\mu_+^x=0.523$, $\mu_+^y=11.4$ that make up the magnification $\mu_+=\mu_+^x\mu_+^y=6.3$.
 In particular $\mu_+^y$  is sensitive to modelling.  A value smaller by, say, a factor 2 will lead in our upcoming analysis
 basically to a microlense distance enhanced by this factor 
 and a mass reduced by it. Such quantitative factors will hardly modify our conclusions.

\section{Models for the microlensing event}

The data of the microlensing event of Figure 2 of Colley and Schild \cite{colley2003rapid} is reproduced in Figure 1. 
Exposed are the $R$-filter data of the A image  (filled circles) minus the interpolated ones of the B image, 
as well as the interpolated A data minus the B data (open circles), with the A (B) data binned at half  (full) hours.
The no-peak -- only-noise explanation is strongly ruled out \cite{colley2003rapid}.
Indeed, it has $\chi^2=80.4$, $\chi^2/\nu=4.02$, so, if the errors are considered as independent Gaussians,
 only with probability $3.3\,10^{-9}$ the cause is a statistical fluke.  
The two data sets are consistent; the 12 data points where the sets overlap have $\chi^2=77.0$,  
this would be a fluke with probability $1.5\,10^{-11}$.
Hence with very high confidence the data represent a physical event. Their sign corresponds to a  lensing event in the B image.
Notice, however, that CS03 do not exclude the possibility that the signal arises due to a dimming event in the A image.

When the center of the microlense crosses the quasar ellipse in the B image at
the angles $\phi_{1,2}$ at times $t _{1,2}$, respectively, we may describe its angular position as

\BEQ 
\br(t) \! =\! \frac{t_2-t}{t_2-t_1}\bx_+(\phi_1)+\frac{t-t_1}{t_2-t_1}\bx_+(\phi_2), \qquad
\EEQ
a linear motion,  and  its angular speed as the constant
 
 \BEQ
 {\bf v}_\ang=\frac{\bx_+(\phi_2)-\bx_+(\phi_1)}{t_2-t_1}.
\EEQ

The inner rim of the accretion disk is a bright, narrow region that carries $c_\isco=25\%$ of the total intensity and the lensing of this part can be detected.
The main part of the disk produces 75\% of the intensity, but does not contribute to the lensing  \cite{schild2005accretion}. 
Each ``infinitesimal'' region ($\phi,\phi+\d\phi$) of the rim (1) produces light rays which in the microlensing event are also magnified by the factors 
(\ref{mupm=}). If both the $\pm$ images contribute, the intensity is enhanced by a factor $\mu=\mu_++\vert\mu_-\vert$, which
combined over the rim results in a magnification

\BEQ
\label{magnie}
 \cE(t)=1+\frac{c_{\isco}}{2\pi}\int_{-\pi}^\pi\d\phi\,\{\mu_+[u(\phi,t)]+\vert\mu_-[u(\phi,t)]\vert-1\}. \hspace{-0.5cm}  \nn\\
\EEQ
Lensing in the B image involves $u(\phi,t)= \vert \bx_+(\phi)-\br(t) \vert  /\theta_E$.
On top of a common, time dependent intensity of both images, 
the data yield an excess $R$-filter magnitude $ m(t)\equiv m_A(t)-m_B(t)-(\bar m_A-\bar m_B)$,
to be modelled by $m(t)=2.5\,\log_{10}\cE(t)$ from (6).

Since the microlensepasses through the view line towards the rim, the impact parameter of the point-lens problem has no meaning 
for this event. 

We now consider various possibilities for the event.

\subsection{A single microlensing event}
In order to improve the modelling successively, 
we start considering a single lensing event, which allows us to make a local straight-line approximation for the ellipse. 
Indeed, if the Einstein angle $\theta_E$ is much less than $\mu_+^y\theta_q$,  the lensing effects occur close to the ellipse, which then
looks as an infinite straight line. For a crossing taking place at the angle $\phi_1$, we approximate 
$\bx(\phi)=\bx(\phi_1)+\bx'(\phi_1)(\phi-\phi_1)$,
while $\br(t)=\bx(\phi_1)+{\bf v}  (t-t_1)$, where only $v_\perp$, the component of ${\bf v}$ perpendicular to $\bx'(\phi_1)$, is relevant.
Setting $A=\vert \bx'(\phi_1)\vert /\theta_E$ and $s=(t-t_1)/t_E$,  where $t_E=\theta_E/v_\perp$ is the typical event duration,
the magnification is obtained by integrating over the line, 

\BEQ 
\label{az}
\hspace{-3mm}\cE(t)&=&1+ \frac{c_\isco}{2\pi  A}\int_{-\infty}^\infty\d \eta\,\left[\mu\left(\! \!\sqrt{s^2+\eta^2}\,\right)-1\right]\nn \\
&=&\hspace{-3mm}1+\frac{c_\isco}{\pi A} \,\frac{1}{\vert s\vert }
\,\left[ (2 + s^2) K\left(\frac{2i}{s}\right) -s^2E\left(\frac{2i}{s}\right) \right].
\EEQ
Here $K(k)$ and $E(k)$ are the complete elliptic integrals of the first and second kinds, respectively.
Near there crossing at $s=0$ there occurs a logarithmic divergency

\BEQ
\label{dE=0}
\cE(s)=1+ \frac{c_\isco}{\pi A}\left(\log\frac{8}{\vert s\vert  }-2+{\cal O}(s^2\log \vert s\vert  )\right),
\EEQ
 a remainder of the $1/u$ singularity of the point-lens case in Eq. (\ref{mupm=}).
It can be cut off by a small but finite thickness of the inner luminous part of the rim.
Since the data points of Fig. 1 are not dense, the peak in $\cE$ will achieve to pierce in between them, 
which prevents a fix of this thickness, hence we neglect it.  The fit yields

\BEQ
 A=29\pm 5, \hspace{1mm}  t_1 =  5.775\pm0.004\, {\rm day} ,  \hspace{1mm}    t_E = 1.0 \pm0.5 \,{\rm  day}. \nn\\
 \EEQ 

\begin{figure}[h]
\centering
\includegraphics[width=7cm]{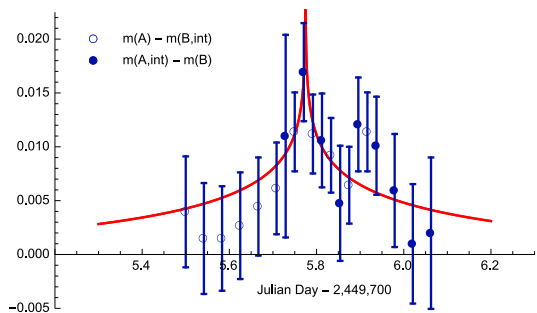}
\caption{Data points: $R$-filter magnitude of the A and B images of the quasar Q0957+561, with respect to each other's interpolations, 
 as function of time in days (Colley and Schild 2003).  Full line:  a single microlensing event in image B.}
 \end{figure}

\noindent
As seen in Fig.1,  this fits well with $\chi^2 =  7.65$. 
The ``too small'' value $\chi^2/\nu=0.45$ (it should equal 1 for Gaussian errors) brings us in the uncommon case of a ``too good fit''.
The same behaviour occurred in the simple Gaussian fit reported by CS03, which led to $\chi^2 = 9.2$ and thus  $\chi^2/\nu=0.51$.
Given that the data represent a physical event with high probability and lie close to the theoretically motivated curve, 
it is very unlikely that we are ``fitting noise'', hence must assume that CS03 have been too conservative 
and overestimated the error bars by at least a factor 1.5 (we will finally argue for a factor 2.5). 
Hence from now on we work with extremely good fits; we shall invoke arguments
other than minimising $\chi^2$ to select between the models.

There is an argument against the present single-peak model, which will also apply to our further models, except the last one.
 The involved logarithmic divergency at the peak
 (whether or not regularised by a small width of the luminous inner rim of the accretion disk) is not expected from the data. 
 Indeed, Fig. 1 of CS03 exposes that not much special is going on with the separate A and B magnitudes.
Moreover, the unbinned data neither show particularly large values that would hint at a narrow peak (R. Schild, private communication). 
Hence we conclude that something else is going on, and that we cannot avoid but looking at fits with even smaller $\chi^2/\nu$. 

The possible logarithmic singularity for the lensing curves has so far not been considered to be a problem in works on the subject, 
supposedly because the luminous disc of the quasar source has finite size, so that the point source--point mass approximation breaks down 
and prevents the divergence. However, this is argument is incorrect, since the divergency stems from the narrowness of the rim,
as exposed explicitly in Eq. (9).

\subsection{Two different microlenses?}

The 20 data points in Figure 1 actually exhibit a double peak structure.
We start with considering that two unrelated microlensing events through the ellipse take place, 
for each of which either the entrance or the exit passing is not documented.

Two microlenses, viz. $m=m_1+m_2$,  fit with  $\chi^2=2.00$  ($\chi^2/\nu=0.14$, $\nu=20-6$) and
$A_1=22 \pm 3$, $t_{1}=5.767 \pm 0.002$ day, $t_{E1}=0.24 \pm 0.07$ day and
$A_2=28\pm  5$, $t_{2}=5.909 \pm 0.002$ day, $t_{E2}=0.14 \pm 0.07$ day.
The fit is presented in Figure 2.

\begin{figure}
\centering
\includegraphics[width=7cm]{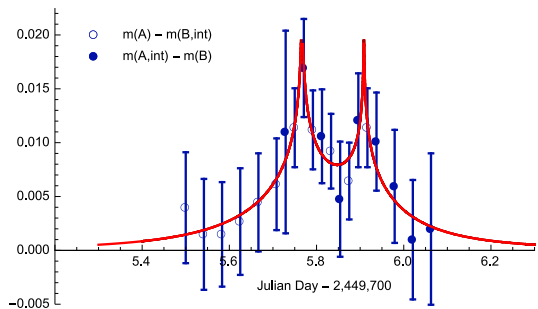}
\caption{Data points as in Fig. 1.  Full line:  two independent microlensing events in image B.
  Except for the logarithmically divergent peaks, the profile coincides within the error bars with the observations.
However, a double peak structure should arise when one microlense moves through the ellipse shaped B image
  of the bright inner rim of the quasar accretion disk.  }
   \end{figure}

 However, the occurrence  of two separate, nearly simultaneous events is rather unlikely, since neither in the previous 4 nights, 
 nor in the 5  full days of monitoring around the world \cite{colley2003around} a lensing event with duration of hours has been observed.
 Hence we consider other options.

\subsection{A microlense intersecting the ellipse twice}

More plausible is that a single microlense enters the quasar ellipse at the time $t_{1}$ of the first peak and exits at time $t_{2}$ 
of the second peak.  
We now have to incorporate the full shape of the ellipse, that is, to employ Eq. (6) with definitions (1) and (5). 
The best fit is very close to the one in Fig. 2, see also its caption, which shows consistency between the approaches.
The present case involves crossing times  $t_1 = 5.768$ $\pm$  $0.002$ day and $t_2 = 5.909\pm 0.004$ day,  
crossing angles $\phi_1 = -2.86\pm 0.02$ and $\phi_2 = 2.75\pm 0.01$. The amplitude $A \equiv\theta_q/\theta_E$ takes the value 
$A = 2.95\pm 0.12$, about 10 times smaller than in previous cases, because now $A$ does not involve a factor $|\bx'(\phi)|\sim\mu_+^y$.
These values are taken at the optimal fit $\beta = 1.54$ and  $\mu_+^y = 23.8$, about twice the above estimate 11.4.
The very good fit,   $\chi^2=2.08$ and $\chi^2/\nu=0.14$ with $\nu=20-5$, would worsen a bit if we would fix $\mu_+^y$ at 11.4.
Ideally, $\beta$, $\mu_+^y$ and  $\mu_+^x $ must
get fixed by global data rather than by the present lensing event, 
but such are not available.

All these details don't really matter much, since, like in previous case, 
this fit suffers from producing two narrow logarithmic divergencies not supported by the data.

\subsection{Partial occultation by a small lensed object}

Partial occultation by the  microlense offers new aspects. Let it have an $R$-filter extinction radius $R_\ext=\theta_\ext d_L$;
light impinging within this radius gets extinct. We  define 
\BEQ
\hspace{-5mm} 
 u_\eE=\frac{\theta_\ext}{\theta_E}-\frac{\theta_E}{\theta_\ext}
 ,\quad
 \frac{\theta_\ext}{\theta_E}=\half(u_\eE+\sqrt{4+u_\eE^2}). 
 \EEQ
For a small lens, that is, in case $\theta_\ext<\theta_E$ and $u_\eE<0$, 
the light,  coming from a position with $u>\vert u_\eE\vert $ has angle $\vert \theta_-(u)\vert <  \theta_\eE$, so when approaching us in the minus image, 
it will impinge on the back side of the microlens, get absorbed and cause partial occultation.
For the quasar lensing, this means that light from regions far enough from the microlense angular positon will be occulted in the minus image.
The plus image, on the other hand, has $\theta_+>\theta_E$ and will not be occulted  \cite{agol2002occultation}. 
The magnification is thus given by (\ref{magnie}), with  $\mu_+(u)$ unmodified and $\mu_-(u)$ replaced  by 0 for $u>\vert u_\eE\vert $. 

This case would keep the logarithmic divergencies upon crossing the quasar ellipse, while the tails of the profile would be smaller
than in Fig. 2. This worsens the fit, even though it remains well acceptable; still the best case is just
the absence of occultation that was treated in section 3.2. Hence we shall no longer consider this possibility.

\subsection{Partial occultation by an extended lensed object}

The absence of logarithmic divergencies in the data can be explained by occultation when $\theta_\ext>\theta_E$, i.e., when $ u_\eE>0$.
In that case the minus image lying at angles $|\theta|<\theta_E$ is always occulted,  
while the plus image is occulted near the central crossing, viz. when $\theta<\theta_\eE$.
The magnification is now given by Eq. (\ref{magnie}), with $\mu_+(u)$ replaced by 0 for $u<u_\eE$ and $\mu_-$ omitted completely.
The logarithmic divergency is thus regularized.

 While we fitted 7 parameters in section 3.2, there are 8 now. The best fit for our model has $\chi^2=2.07$. 
The errors in  $\mu_+^y$ and $\beta$ may again be set to zero,  since these parameters should be determined from other observations.  
Hence we count $\nu=20-6$ degrees of freedom, which leads to $\chi^2/\nu=0.148$. 
Also this lies below the Gaussian value 1,  so we again conclude that the errors 
have been taken too conservatively by Colley and Schild \cite{colley2003rapid}, and should probably be about 2.5 times smaller.
After repairing by hand for this factor,  we have artificially constructed a data set with $\chi^2/\nu\approx1.0$. The best fit is then:

 \BEA
&& t_1= 5.770\pm0.001\,{\rm day},\quad t_2= 5.917  \pm 0.002\,{\rm day} \nn\\
&& A= 1.08 \pm 0.01,\qquad \quad\quad u_\ext= 0.10\pm 0.02 \nn\\
&& \phi_1=3.04 \pm 0.03, \qquad\qquad  \phi_2= 0.64 \pm 0.01. \EEA
(One may well argue that the errors in the data should not be diminished, so that the errors in Eq. (12) should be remultiplied by the factor 2.5.)
The value for $u_\ext$ is small, but not unrealistic.
The optimal microlense trajectory is plotted in Fig. 3. The fit is presented in Fig. 4 with error bars of the data points  artificially reduced by the factor 2.5.

\begin{figure}
\centering
\includegraphics[width=7cm]{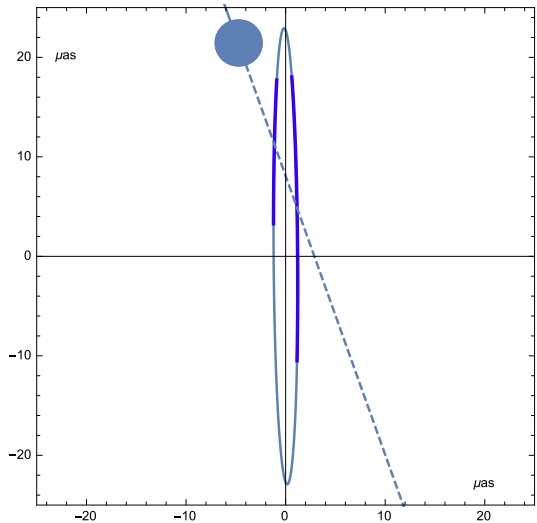}
\caption{In the B (and also A) image the bright quasar inner rim is viewed as a nearly perfect ellipse perpendicular to the quasar-lens-Sun plane,
indicated by the horizontal axis. The rim cannot be resolved optically, but the microlense does so and produces a doubly peaked lensing signal.
The fat parts of the ellipse indicate the error bars on the crossing points.
When the extinction radius of the lens exceeds its Einstein radius, partial occultation takes place which regularises infinite lensing peaks.  }
\label{LensInOut}
\end{figure}

\begin{figure}
\label{muLtraj}
\centering
\includegraphics[width=7cm]{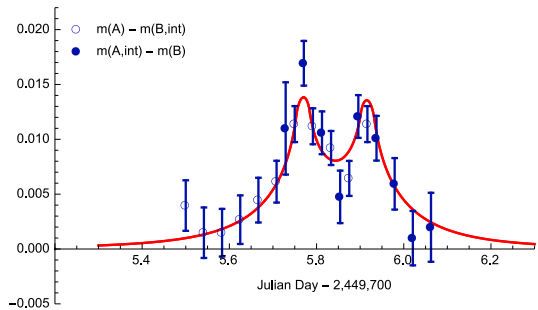}
\caption{Lensing with occultation of the near part of the quasar sightline truncates the divergencies of the  peaks. 
 Data as in Fig. 1, with error bars artificially reduced by a factor $2.5$ so as to achieve $\chi^2/\nu\approx1$.}
 \end{figure}

\section{Physical interpretation of the  models}

 Let us interpret these cases in terms of the Einstein radius, mass, distance and, where possible, the light extinction radius and temperature of the microlense.

1) The  single microlense of section 3.1 has $\theta_E=|\bx_+'(\phi_1)|/A \sim \mu_+^y\theta_q/A\sim 0.92\,\mu$as.
Assuming a typical speed $v_\perp\sim 220$ km/s, we get a Mpc distance and Earth scale mass,

\BEQ 
\hspace{-3mm} 
d_L=\frac{v_\perp t_E }{\theta_E}\sim 144 \, {\rm kpc}, \quad
M=\frac{c^2d_L\, \theta_E^2}{4G}\sim 5.0\,M_\oplus. 
\EEQ
These values are not unreasonable and refer to a microlense in the Galactic halo.
But, as remarked, a sharp lensing peak is not exposed by the data.

The speed $v_\perp\sim 220$ km/s is consistent with the microlense lying in the Galaxy, and this will remain the case
for the more precise models to be considered next.

2) For the two microlenses of section 3.2 and $v_\perp= 220$ km/s, we get Einstein radii, distances and masses 

\BEQ 
\theta_E&=&1.2\,\mu{\rm as},\quad d_L=25\,{\rm kpc},\quad M=1.5\,M_\oplus, \quad {\rm day\,\, 5.77} , \nn\\
\hspace{-3mm} 
\theta_E&=&1.0\,\mu{\rm as},\quad d_L=18\,{\rm kpc},\quad M=0.7\,M_\oplus, \quad {\rm day\, \,5.92} . \nn\\
\EEQ
These events appear to happen well within the Galaxy.
But, as stated above, the chance for the occurrence of two independent microlenses in the same night is slim.

3) In the case of section 3.3, a single microlense going into and out of the sight of the quasar ellipse, one has
$\theta_E=\theta_q/A=0.81\mu$as. The angular speed is $v_{\rm ang}= 8.2\,\, \mu{\rm as}\,/{\rm day}.$
A speed $v_{\rm rot}=220$ km/s sets the distance and mass as

\BEQ
\hspace{-3mm} 
d_L=\frac{v_{\rm rot}}{v_{\rm ang}}=16\,\kpc, \quad M=\frac{c^2d_L\,\theta_E^2}{4G}= 0.41\,M_\oplus.
\EEQ
The values are consistent with previous cases, but the model still suffers from the logarithmic divergencies.

4) Partial occultation by a small object  leads us back to the no-occultation case 3).

5) Lastly, we consider the case of section 3.5, a large lens causing partial occultation during the passages of the rim, thereby
smoothening the logarithmic divergencies.
One gets $\theta_E=\theta_q/A=2.2\,\mu$as  and $v_{\rm ang}=45\,\mu$as day$^{-1}$.
 Hence its distance and mass are

\BEQ
d_L=\frac{v_{\rm rot}}{v_{\rm ang}}=2.8\, {\rm kpc}, \qquad M=0.56\,M_\oplus,
\EEQ
so it lies nearby in the Galaxy, having half of the Earth mass.  With $u_\ext$ from (11),
the physical Einstein radius  $R_E=\theta_Ed_L=1.3\, R_\odot$ allows to fix the $R$-filter extinction radius,

\BEQ
R_\ext=\frac{1}{2}(u_\ext+\sqrt{4+u_\ext^2})R_E=1.4\,R_\odot,\hspace{2mm} \frac{R_\ext}{d_L}=2.4\mu{\rm as}.
\nn\\
\EEQ
Assuming that the lensing object is an isothermal, primordial gas cloud with radius not much larger than this, 
we take from the introduction the estimate for the temperature

\BEQ
T= \frac{GM\mu m_N}{2k_B R_\ext}=17\,{\rm K}.
\EEQ
This lies in the range of many observations and above the $\sim13$ K transition from gas to liquid,
as it should, because as a liquid or solid the lens would be much more compact.

The numbers in this section depend essentially linearly on our value $\mu_+^y=11.4$.
To consider an extreme case, we note that a value of $\mu_+^y$ smaller by a factor 2 leads basically to a microlense distance enhanced by this factor 
and a mass reduced by it, so we remain in the ballpark of dozens of parsec distances and Earth scale masses.
While the factor would cancel from its physical Einstein radius $R_E$, so that it would not change much, the temperature would be reduced by basically this factor.
Though such a $T<14$ K temperature is in conflict with the physics of an extended cloud, the 14 K case may well remain possible within our error bars.

\section{On the possibility of occultation in microlensing of LMC stars}

The EROS and MACHO teams have ruled out compact objects --  such as black holes, brown dwarfs and e.g. Jupiter-like gas planets  --
as putative MACHOs that make up the Galactic dark matter,
but the conclusion by the MACHO team has been questioned recently by Hawkins \cite{hawkins2015new},
because the Galactic halo appears to be less heavy than it was assumed a decade ago.
Hence let us reconsider the case of lensing of the Large Magellanic Cloud (LMC).

We restrict ourselves to stars small enough for the point-lens approximation to apply.
For an isothermal cloud positioned at distance $d_L\equiv xd_\LMC$ ($0<x<1$) towards the LMC 
one may derive  that  ${R_\ext}/{R_E}=\sqrt{{M}/{M_c}}$, where

\BEQ 
\hspace{-7mm} 
M_c=\frac{16 (1-x)d_{L}\sigma_\gas^4}{r_\ei^2Gc^2}
=1.7\,\frac{x(1-x)}{r_\ei^2}\left(\frac{T}{15\, \K}\right)^2M_\oplus,
\EEQ

\noindent\noindent
with $r_\ei=R_\ext/R_\iso\ge 1$. For $M>M_c$ the situation is the one underlying the analysis of section 3.4:
the minus image is always occulted, while the plus image is occulted upon near passage (for $u<u_\eE$), 
whence the divergency of the lensing {\it peak} is truncated and replaced by an occultation {\it dip} in the middle,
i.e., a signal with two peaks.


\section{Connection to the theory of gravitational hydrodynamics}

Below redshift $z=160$ there are not enough photons to keep baryons at the photon temperature; instead the velocities diminish linearly in the 
scale factor, so that their temperature goes quadratically, $T_b=(1+z)^2\,  0.017$ K \cite{loeb2001reionization}. 
Hence at $z\simeq 28$ all hydrogen passes through the triple point temperature 13.8 K,
allowing it to condense (liquify or even freeze) and form stars and galaxies from the age of 110 Myr on.

With neither WIMPs nor supersymmetry observed, it may pay off to consider alternatives for the theory of $\Lambda$ cold dark matter.
Gravitational hydrodynamics (GHD) presupposes nonlinear structure formation and puts forward that after the recombination, 
all gas breaks up in Jeans clumps of millions of solar masses, which themselves fragment into primordial cloudlets of about Earth mass, called micro brown dwarfs (\mBDs);
this turns the Jeans clumps into Jeans clusters of \mBDs \, \cite{Gibson1996Turbulence,nieuwenhuizen2010micro}.
Millions of Jeans clusters (JCs) then must embody the Galactic dark matter; this bold picture obviously softens the missing baryon problem.
Till now, there have not been direct observations that support the existence of these Jeans clusters or \mBDs;
however, it appears that we can consider the discussed microlensewith its Earth-scale mass as a \mBD \ candidate.

GHD explains several problems at the Galactic scale, such as:  the flattening of rotation curves;  the origin of young star clusters in ``tidal tails'' of galaxy mergers; 
and the iron core problem \cite{nieuwenhuizen2010micro} ;  the Helium-3 problem and the wide-binaries problem \cite{nieuwenhuizen2011explanation};
and also the last-parsec problem, the fast growth of super massive black holes; a maximum in the star formation rate; and
the relation between the mass of the central black hole, the one of the bulge, and the number of globular star clusters
\cite{nieuwenhuizen2012model}. 

GHD has a nonlinear, top down structure formation with a dark age of 110 millions of years only 
\cite{Gibson1996Turbulence,nieuwenhuizen2009gravitational,schild2011turbulent}.
The nonlinear structures would not be washed out by free streaming  neutrinos.
Hence eV-scale masses of active and possibly also sterile neutrinos is not ruled out in GHD.

GHD thus welcomes very early galaxies, like the one reported at redshift $z=11.9$ \cite{ellis2012abundance}.
It also explains why high redshift galaxies may already look regular and that 
nearby galaxies appear more regular than expected from standard cosmology  \cite{disney2008galaxies}.
Unfortunately, the GHD  theory has not been shown to explain the cosmic microwave background spectrum and the matter power specturm
in structure formation, while Big Bang Nucleosynthesis should likely involve chemical potentials for neutrinos \cite{nieuwenhuizen2009nonrelativistic}.

Cold $\ge$15 K gas clouds are abundant and often related to spinning dust.
The cause for such a common lowest temperature despite very different environments is unknown, but \mBDs \ provide an explanation. 
Being in principle primordial H-He clouds that tend to cool down to 2.7 K,
they need to pass the first order gas-to-liquid/ice transition line that starts from the triple point at $T=13.8$ K and $p=7.04$ kPa, 
and goes to $T\sim 13$ K at $p=0$. 
There are two cases: colder clouds  which liquify and squeeze, difficult to observe, and those heated by stars keeping the temperature
above $\sim13$ K and keeping their solar size.
The resulting gas cloud with unresolved density concentrations, the \mBDs, then comprises much more mass than estimated currently.
The H clouds themselves having a 15 K temperature allows to thermostate the dust temperature at 15 K,
without the need of assistance by stellar radiation.
Evidently, the hidden baryonic mass would contribute to the Galactic ``dark matter" budget, if not being all  of it.

Let us return to the potential microlenses in the Galaxy.  Our  MACHO should be an \mBD \  belonging to
a Jeans cluster that lies in front of the Twin Quasar; unfortunately no Herschel picture is available to confirm this.

The optical depth for the partial occultation case is calculated as follows. 
Let the Jeans cluster have a mass $M_\jc\sim 10^5M_\odot$ 
and radius $R_\jc\sim 2$ pc, so that it consists of  $N=3\,10^9\bar M/M_\oplus$ of  \mBDs \ with average mass $\bar M$.
The angular area of an JC is $\Omega_\jc=\pi R_\jc^2d_\muL^{-2}$ and the one of the Twin Quasar ellipse $\Omega_{\rm tq}=\pi \mu_+ \theta_q^2$. 
The typical optical depth is
$\tau=N\Omega_{\rm tq}/\Omega_\jc=2 \,10^{-4}(M_\jc/10^5M_\odot)(M_\oplus/M_\bd)({\rm pc}/R_\jc)^2B_{\it f}$.
The boost factor $B_{\it f} \lesssim 10$ arises because the centre of the  postulated Jeans cluster may lie in front of the B image.
The cadence $\tau v_\ang/\theta_E$ $=1.6 B_{\it f}(M_\jc/10^5M_\odot)(M_\oplus/M_\bd)\times $ $({\rm pc}/R_\jc)^2$ yr$^{-1}$
can be as large as once per month. 

For the 5 observation nights in 1996/1998 and the 10 full days in 2000/2001, 
we would estimate the number of events to be maximally  $12.5$ day $\times$ 1/month = 0.4, which is very reasonable given
that one event was indeed observed, the one analysed above.
In conclusion, there appears to be a consistency between the properties of the lensing event and the theory of gravitational hydrodynamics.

\section{Conclusion}

We analyse five quasar microlensing models for the Colley - Schild (CS03) lensing event in the Twin Quasar \cite{colley2003rapid},
which leads to a consistent picture of an event with Einstein radius of 1 $\mu$as, due to a lens with Earth-scale mass 
and located inside the Galaxy. 
We face the paradox of too small error bars ($\chi^2/\nu\sim0.15$) together with a very small ($\sim 10^{-9}$) probability
that the data do not represent a physical event. 
Like in the Gaussian fit by CS03, we take the position that the event is real and that  error bars have been overestimated.

The peaks in the data are related to passing into and out of the quasar ellipse in the B-image.
Narrow divergencies, not exposed by the data, are suppressed in the theory by allowing for occultation, 
which can be caused by a dense, primordial gas cloud of Solar size, large enough to occult a substantial part of the light paths.
Viewed as isothermal, its estimated temperature of 17 K lies above the phase transition near 13 K, 
below which the cloud would liquify and be much smaller.
It is remarkable that the resulting mass of about half that of the Earth, falls in the range predicted from Gravitational Hydrodynamics
and is comparable to the estimated $\sim 3\,M_\oplus$ mass of a population of MACHOs in the lensing galaxy of the same  Twin Quasar
\cite{schild1996microlensing}. Such clouds should not evaporate in a Hubble time \cite{deRujula1992nature}.

For microlensing the Twin Quasar is more ideal than point sources, since we observe the inner rim of the accretion disk as an ellipse, which is intersected twice. 
Hence when a microlense in the Galaxy passes through, it will have a zero impact parameter and two peaks in the intensity.
Even with moderate precision data, we have determined separately the mass, distance and size of the microlens.
Hence it would be interesting to have continuous data of the Twin Quasar's images, which allows both for a good determination of the
present time delay between the two images, and to filter out possible microlensing events.

When the MACHOs are not compact objects such as black holes, but cold, solar-size gas clouds, occultation can occur during the lensing event.  
For lensing towards the Magellanic Clouds, this was first put forward in  \cite{nieuwenhuizen2010micro} and then further considered in \cite{nieuwenhuizen2012herschel}. 
The theory of Gravitational Hydrodynamics (GHD) predicts that the MACHOs are embedded in Jeans clusters, 
and may explain a part of the Galactic dark matter, the other part coming from cold or warm dark matter,
or be all the dark matter if those forms of matter do not exist.
The baryonic dark matter of the Galaxy would then be composed of Jeans clusters with mass of millions of solar masses,
 fragmented in cloudlets of Earth mass which, if warmer than 13K, have solar size, just like the object that emerged above as the most likely quasar microlens.
The non-baryonic dark matter of galaxy clusters may emerge from neutrinos with eV-scale mass
\cite{nieuwenhuizen2009nonrelativistic,nieuwenhuizen2011prediction,nieuwenhuizen2013observations}.
The latest work in this series finds a perfect match for the A1689 cluster and the cosmic dark matter amount \cite{nieuwenhuizen2016dirac}.

While in the theory of linear structure formation neutrinos would wash out relevant parts of the structures,
they may not achieve that for the GHD nonlinear structure formation, and hence reappear as acceptable dark matter candidates.

To settle these fundamental questions, we propose to pay further attention to the possibility of occultation in microlensing.
This will be possible in upcoming full time monitoring of the cosmos.

\vspace{3mm}

{\it Acknowledgments}: We are grateful for discussion with 
Rudy Schild and Peter Keefe.

\vspace{3mm}

keywords: {microlensing, Twin Quasar, MACHO, occultation}



\begin{thebibliography}{[10]}

\bibitem{shull2012baryon}
 \textsc{J.\,M. Shull},  \textsc{B.\,D. Smith},  and  \textsc{C.\,W. Danforth}
  \jr{The Astrophysical Journal} \textbf{759}(1), 23 (2012).


\bibitem{schild1996microlensing}
 \textsc{R.\,E. Schild} \jr{The Astrophysical Journal} \textbf{464}, 125
  (1996).


\bibitem{renault1998search}
 \textsc{C.~Renault},  \textsc{E.~Aubourg},  \textsc{P.~Bareyre} 
\etal{} \jr{Astronomy and Astrophysics} \textbf{329},  522--537 (1998).


\bibitem{tisserand2007limits}
 \textsc{P.~Tisserand},  \textsc{L.~Le~Guillou},  \textsc{C.~Afonso},
 \etal{} \jr{Astronomy \& Astrophysics} \textbf{469}(2),  387--404 (2007).


\bibitem{alcock1998eros}
 \textsc{C.~Alcock},  \textsc{R.~Allsman},  \textsc{D.~Alves},
\etal{} \jr{The Astrophysical Journal Letters}  \textbf{499}(1), L9 (1998).


\bibitem{alcock2000macho}
 \textsc{C.~Alcock},  \textsc{R.~Allsman},  \textsc{D.\,R. Alves},
\etal{} \jr{The Astrophysical Journal} \textbf{542}(1),  281 (2000).


\bibitem{microlensing2011unbound}
  \textsc{O.\,G.\,L.\,E.\,O. Collaboration} \etal{} \jr{Nature}
  \textbf{473}(7347), 349--352 (2011).


\bibitem{clanton2016constraining}
 \textsc{C.~Clanton} and  \textsc{B.\,S. Gaudi} \jr{arXiv preprint
  arXiv:1609.04010} (2016).


\bibitem{ban2016microlensing}
 \textsc{M.~Ban},  \textsc{E.~Kerins},  and  \textsc{A.~Robin} \jr{arXiv
  preprint arXiv:1606.06945} (2016).


\bibitem{hawkins2015new}
 \textsc{M.~Hawkins} \jr{Astronomy \& Astrophysics} \textbf{575}, A107 (2015).


\bibitem{ade2015planck}
 \textsc{P.\,A. Ade},  \textsc{N.~Aghanim},  \textsc{M.~Arnaud},
 \etal{} \jr{Astronomy \& Astrophysics} \textbf{582}, A28  (2015).


\bibitem{juvela2011galactic}
 \textsc{M.~Juvela},  \textsc{I.~Ristorcelli},  \textsc{V.\,M. Pelkonen},
\etal{} \jr{Astronomy \& Astrophysics} \textbf{527},  A111 (2011).


\bibitem{juvela2012galactic}
 \textsc{M.~Juvela},  \textsc{I.~Ristorcelli},  \textsc{L.~Pagani},
\etal{} \jr{Astronomy \& Astrophysics} \textbf{541}, A12  (2012).


\bibitem{colley2003around}
 \textsc{W.\,N. Colley},  \textsc{R.\,E. Schild},  \textsc{C.~Abajas},
\etal{} \jr{The Astrophysical Journal} \textbf{587}(1), 71  (2003).


\bibitem{colley2003rapid}
 \textsc{W.\,N. Colley} and  \textsc{R.\,E. Schild} \jr{The Astrophysical
  Journal} \textbf{594}(1), 97 (2003).


\bibitem{deRujula1992nature}
 \textsc{A.~De~R{\'u}jula},  \textsc{P.~Jetzer},  and  \textsc{E.~Mass{\'o}}
  \jr{Astronomy and Astrophysics} \textbf{254}, 99 (1992).


\bibitem{schild2005accretion}
 \textsc{R.\,E. Schild} \jr{The Astronomical Journal} \textbf{129}(3), 1225
  (2005).


\bibitem{hinshaw2013nine}
 \textsc{G.~Hinshaw},  \textsc{D.~Larson},  \textsc{E.~Komatsu},
\etal{} \jr{The Astrophysical Journal Supplement Series}  \textbf{208}(2), 19 (2013).


\bibitem{riess20113}
 \textsc{A.\,G. Riess},  \textsc{L.~Macri},  \textsc{S.~Casertano},
\jr{The  Astrophysical Journal} \textbf{730}(2), 119 (2011).


\bibitem{fadely2010improved}
 \textsc{R.~Fadely},  \textsc{C.\,R. Keeton},  \textsc{R.~Nakajima},  and
  \textsc{G.\,M. Bernstein} \jr{The Astrophysical Journal} \textbf{711}(1), 246
  (2010).


\bibitem{agol2002occultation}
 \textsc{E.~Agol} \jr{The Astrophysical Journal} \textbf{579}(1), 430 (2002).


\bibitem{loeb2001reionization}
 \textsc{A.~Loeb} and  \textsc{R.~Barkana} \jr{Annual Review of Astronomy and
  Astrophysics} \textbf{39}, 19--66 (2001).


\bibitem{Gibson1996Turbulence}
 \textsc{C.\,H. Gibson} \jr{Applied Mechanics Reviews} \textbf{49}, 299--315
  (1996).


\bibitem{nieuwenhuizen2010micro}
 \textsc{T.\,M. Nieuwenhuizen},  \textsc{R.\,E. Schild},  and  \textsc{C.\,H.
  Gibson} \jr{arXiv preprint arXiv:1011.2530} (2010).


\bibitem{nieuwenhuizen2011explanation}
 \textsc{T.\,M. Nieuwenhuizen} \jr{arXiv preprint arXiv:1103.6284} (2011).


\bibitem{nieuwenhuizen2012model}
 \textsc{T.\,M. Nieuwenhuizen} \jr{EPL (Europhysics Letters)} \textbf{97}(3),
  39001 (2012).


\bibitem{nieuwenhuizen2009gravitational}
 \textsc{T.\,M. Nieuwenhuizen},  \textsc{C.\,H. Gibson},  and  \textsc{R.\,E.
  Schild} \jr{EPL (Europhysics Letters)} \textbf{88}(4), 49001 (2009).


\bibitem{schild2011turbulent}
 \textsc{R.\,E. Schild} and  \textsc{C.\,H. Gibson} \jr{arXiv preprint
  arXiv:1105.1539} (2011).


\bibitem{ellis2012abundance}
 \textsc{R.\,S. Ellis},  \textsc{R.\,J. McLure},  \textsc{J.\,S. Dunlop}
\etal{} \jr{The Astrophysical Journal Letters}
  \textbf{763}(1), L7 (2012).


\bibitem{disney2008galaxies}
 \textsc{M.~Disney},  \textsc{J.~Romano},  \textsc{D.~Garcia-Appadoo}
 \etal{} \jr{Nature} \textbf{455}(7216), 1082--1084 (2008).


\bibitem{nieuwenhuizen2009nonrelativistic}
 \textsc{T.\,M. Nieuwenhuizen} \jr{EPL (Europhysics Letters)} \textbf{86}(5),
  59001 (2009).


\bibitem{nieuwenhuizen2012herschel}
 \textsc{T.\,M. Nieuwenhuizen},  \textsc{E.\,F. van Heusden},  and
  \textsc{M.\,T. Liska} \jr{Physica Scripta} \textbf{2012}(T151), 014085
  (2012).


\bibitem{nieuwenhuizen2011prediction}
 \textsc{T.\,M. Nieuwenhuizen} and  \textsc{A.~Morandi} \jr{arXiv preprint
  arXiv:1103.6270} (2011).


\bibitem{nieuwenhuizen2013observations}
 \textsc{T.\,M. Nieuwenhuizen} and  \textsc{A.~Morandi} \jr{Monthly Notices of
  the Royal Astronomical Society} p.\,stt1216 (2013).


\othercit
\bibitem{nieuwenhuizen2016dirac}
 \textsc{T.\,M. Nieuwenhuizen},
 Journal of Physics: Conference Series,  (2016),  p.\,012022.


\end{thebibliography}

\providecommand{\WileyBibTextsc}{}
\let\textsc\WileyBibTextsc
\providecommand{\othercit}{}
\providecommand{\jr}[1]{#1}
\providecommand{\etal}{~et~al.}

\end{document}